\documentclass[10pt,prl,twocolumn,showpacs]{revtex4}

\usepackage{graphicx}%
\usepackage{dcolumn}
\usepackage{amsmath}
\usepackage{epsfig}

\makeatletter
\def\btt#1{\texttt{\@backslashchar#1}}%
\DeclareRobustCommand\bblash{\btt{\@backslashchar}}%
\makeatother

\begin{document}

\preprint{HEP/123-qed}

\title[Model for Gravitational Interaction between Dark Matter and Baryons]{Model for Gravitational Interaction between Dark Matter and Baryons}

\author{Federico Piazza$^{1,2}$ and Christian Marinoni$^3$}
\affiliation{$^1$Dipartimento di Fisica, Universit\`a di Milano Bicocca, Piazza delle Scienze 3, I-20126 Milan, Italy\\
$^2$Institute of Cosmology and Gravitation, University of Portsmouth, Portsmouth PO1 2EG, UK \\
$^3$ Laboratoire d'Astrophysique de Marseille,  Traverse du Siphon B.P.8, 13376 Marseille, France}

\date{\today}

\begin{abstract}
We propose a phenomenological model where the gravitational interaction between dark matter and baryons
is suppressed on small, subgalactic scales. We describe the  gravitational force  by adding    
a Yukawa  contribution to the standard Newtonian potential and show that this interaction scheme 
is effectively suggested by the available observations of the inner rotation curves of small mass galaxies.
Besides helping in interpreting the cuspy profile of dark matter halos observed in N-body simulations, 
this potential regulates the quantity  of baryons within halos of different masses.
\end{abstract}

\pacs{98.35.+d, 98.62.Gq, 95.30.Sf}

\maketitle
 Lacking   evidence of  direct detection, the  presence, nature and quantity of dark matter (DM)   
 must be in general  inferred from  the kinematic and distribution  properties of baryons. 
 A model based on the growth of small fluctuations 
 through gravitational instability in a universe dominated by  cold dark matter (CDM) provides a satisfactory fit 
 to a wide range of properties of visible structures  on scales $\gtrsim 1$ h$^{-1}$Mpc  \cite{PEA}. 
 Recently, however, some apparent inconsistencies in relating DM and baryon properties  on subgalactic scales
 have  been highlighted, prompting the investigation of radical  models to explain the discrepancy.

 Low surface brightness (LSB) galaxies are dominated by dark
 matter \cite{BLO} whose distribution can be traced  by the rotation properties of a dynamically 
 unimportant stellar component. Therefore, these galaxies provide the best laboratories where to test the consistency of the 
 CDM scenario.  Recent high-resolution studies of the inner ($\lesssim$ 5 kpc) rotation curves of LSB galaxies
 suggest that the dark matter is  distributed in spherical halos with  nearly constant density cores ($\lesssim$ 1 kpc) 
\cite{MCG,BL2,MAR}.
These observations seem  to collide with the predictions of CDM
 models which produce halos with compact
 density cusps and steep, universal  mass-density profiles $\rho \sim  r^{-\beta}$ with $\beta \sim$ 1-1.5   
(e.g. \cite{MOO,NFW}).
This disagreement must now be taken seriously since  it is not easily 
 interpreted in terms of observational uncertainties \cite{BBG} or 
  artificial resolution effects in the CDM  simulations \cite{GHI}.

 A parallel problem  is that hierarchical clustering theories   with scale-invariant primordial perturbations
 show significantly more virialized objects of dwarf-galaxy mass in a typical galactic halo 
than are observed around the Milky Way  \cite{KWG}.  
 Possible solutions to this  conflict   range from non-standard models 
 of inflation  \cite{KAM} or  models in which the Universe is
 dominated by warm, self-interacting or annihilating DM (e.g. \cite{SPE}),
 to hydrodynamical  mechanisms that  prevent accreting baryons to cool into stars (e.g. \cite{BAR}). 

 In the absence of an obvious astrophysical mechanism, 
 the purpose of this note is to investigate if the various mismatches between observations  
 and CDM predictions, which however appear to be characterized  by a common physical length at subgalactic 
 scales, reflect a more fundamental theoretical  problem concerning the nature of the gravitational 
attraction between exotic (dark)  and standard (visible)  matter. 

 Occasionally, modifications of gravity have been proposed on scales 
 larger than those investigated here in order to  describe the flatness of the rotation curves without 
the need of dark matter  \cite{FIN} (see \cite{SCO} for a review). 
 The picture that we propose, on the contrary,  aims to modify   
neither the successful DM paradigm nor the standard baryonic physics at any  scale length. 
We speculate 
that, in the Newtonian limit of approximation 
(small curvatures and small velocities in Planck units), 
the gravitational interaction in the \emph{mixed}, 
dark--visible sector (i.e. that between baryonic and non baryonic particles), 
 is dampened  on small ($\sim $ kpc) scales. 

We parameterize a general small-scale modification of gravity in the mixed sector, by adding to the standard
``$1/r$'' potential a Yukawa contribution which is active only between visible and dark matter particles.
This choice is motivated by the known short-range character of the Yukawa interaction
and by its simple analytical properties in both position and momentum space.

Accordingly, we  rewrite the total gravitational potential  acting on the baryons as 
$\phi = \phi_N + \alpha \, \phi_Y $, where $\phi_N$ is the usual Newtonian potential, 
\begin{equation}\label{11}
\nabla^2 \phi_N ~ = ~ 4 \pi G \, (\rho_D + \rho_{B}),
\end{equation}
and where, weighted by the strength parameter $\alpha$, $\phi_Y$ is the Yukawa contribution: 
\begin{equation}\label{22}
(\nabla^2 - \lambda^{-2})\, \phi_Y = 4 \pi G \, \rho_D.
\end{equation}
The parameter $\lambda$ is the typical length scale above which $\phi_Y$ dies out exponentially.
Note that the Yukawa contribution  to the total gravitational potential ``felt'' by baryons 
is sourced by the  DM energy density $\rho_D$ only.

By inverting eq. \eqref{22} the familiar exponential behavior of the Yukawa potential is recovered.
Neglecting the contribution of the baryons in eq. \eqref{11}, 
the gravitational potential $\phi({\bf x})$ felt by a visible test particle 
in the presence of a distribution of DM $\rho_D({\bf x'})$ amounts to
\begin{equation} \label{1}
\phi({\bf x})~=~ - G\int d^3x'\rho_D({\bf x'})\frac{1+
\alpha \, e^{- |{\bf x} - {\bf x'}|/\lambda}}{|{\bf x}- {\bf x'}|} \ .
\end{equation}
For the critical value $\alpha=-1$ the
gravitational attraction between two point particles of different species 
does not diverge and is maximally suppressed in the small distance limit.
We have tested our hypothesis of gravitational suppression
by fitting the parameters of this Yukawa model  using the rotation 
curves of LSB galaxies.

We perform  the integration \eqref{1} in the case of a spherically symmetric distribution of   
DM and we find that the contribution of a spherical shell of radius $r'$ and thickness $dr'$
to the Yukawa part of the potential is given by
\begin{equation*}
\frac{d\phi_Y(r)}{2 \pi G}= -\frac{ \lambda}{r} \left(e^{-|r-r'|/\lambda}  -  e^{-(r+r')/\lambda} \right) r'
\rho_D(r') dr' .
\end{equation*}
Note that since Gauss theorem does not hold for $\phi_Y$ (see eq. \eqref{22})
a spherical shell at $r'$ generally does contribute 
to the net force at some internal point $r < r'$. 

A general prediction of CDM models is that dark matter halos should 
have a steep central cusp, i.e. their density profiles have pure \emph{power-law forms}. 
By introducing a rescaled dimensionless radius $x \equiv r/\lambda$,
we thus focus on a density profile of the type 
$\rho_D(x) = \rho_0\, x^{-\beta}$.  
In this relevant case the integral \eqref{1} can be solved analytically 
in terms of confluent hypergeometric functions of the first kind 
$\Phi (\alpha;\, b;\, z)$. The velocity curve, $v^2 = r \, |d\phi/d r|$, is given by 
\begin{equation} \label{6}
\frac{v^2(x)}{4 \pi G \rho_0 \lambda^2} = \frac{x^{2-\beta}}{3-\beta} + \frac{\alpha}{2(2-\beta)} \left[
F(x, \beta)+ F(-x, \beta) \right]\, ,
\end{equation}
where 
\begin{equation*}
F(x, \beta) = \left(1 - \frac{1}{x}\right) \left[|x|^{2-\beta} \Phi(1;3-\beta;x)  - \Gamma(3-\beta) e^x \right].
\end{equation*}

At large radii ($x \gg 1, r \gg \lambda$), the velocity curve \eqref{6} flattens down
to the standard Newtonian behavior, $v \propto x^{(2-\beta)/2}$. In the inner regions, however, 
the Yukawa contribution is efficient and the curve may be approximated  by a different power-law;
in this case the Newtonian exponent for the velocity gets a correction that we 
have estimated as $- \alpha[1 + \beta(1-\alpha)]/11$.
In the case of a NFW \cite{NFW} inner profile ($\beta = 1$), eq. \eqref{6} reduces to   
\begin{equation*} 
\frac{v^2(x)}{4 \pi G \rho_0 \lambda^2} ~=~ \frac{x}{2} + \frac{\alpha}{x}[1-e^{-x}(1+x)], \quad (\beta = 1) .
\end{equation*}

\begin{figure}
 \includegraphics[scale=0.54]{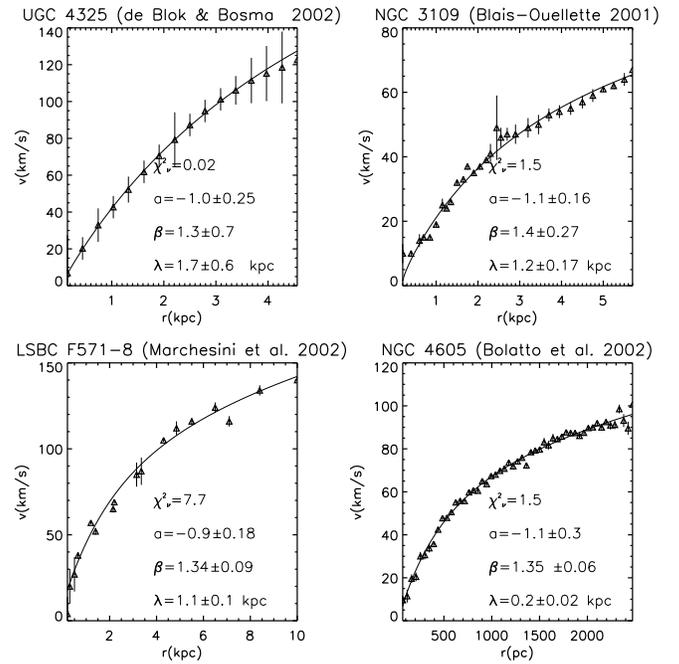}%
 \caption{
Velocity  models derived by fitting the high resolution rotation curves  of LSBC F571-8 \cite{MAR}, NGC 4325 \cite{BL2}, NGC 3109 \cite{BLA}, and NGC 4605 \cite{BSL}  with eq. 5. The fit is performed
assuming no dynamical contribution from the disk (minimal disk hypothesis). 
The best fitting parameters for each galaxy are shown in the corresponding panel. 
Errorbars represent 1$\sigma$ uncertainties. 
\label{uno}}
\end{figure}

We have investigated the consistency and  the universality of this gravitational paradigm, by 
fixing the model parameters by means of  the observed (but theoretically unexplained)  kinematics 
of low surface brightness  galaxies. 
To this purpose we have used high spatial  resolution H$_{\alpha}$ rotation curves  measured and 
reduced by different authors \cite{BL2, BLA, MAR, BSL}  
according to independent observational strategies and technics (see fig. 1). 

We  have assumed that the LSB halo is the dominant mass component and we have not 
attempted to model and remove the contribution from the galactic disk to the rotation, which is taken to be 
negligible ({\em minimal disk} hypothesis, i.e. the mass-to-light (M/L) ratio of the material in the 
disk is M/L $\sim 0$, e.g. \cite{BSL}).  Our results may be described as follow:\\
 {\it a)} the parameter $\alpha$ is remarkably stable for all the objects considered.
Moreover, its mean value $\alpha=-1 \pm 0.1$ physically suggests that, 
independently of the phenomenologically adopted Yukawa model, 
the effective force between baryons and DM tends to {\em vanish} 
on scales smaller than some characteristic length $\lambda$; \\ 
{\it b)} the best fitting value for  $\beta$ ($1.35 \pm 0.05$), the inner slope of the  halo mass  
density universal  
profile, is in excellent agreement  with the CDM calculations by \cite{NFW, GHI}. 
In other terms the observed  kinematics of visible matter in dwarf galaxies  is dynamically  induced  by the 
dark matter clumps  predicted by numerical simulations. As it is well known, 
 a pure Newtonian potential fails to reconcile observations with theoretical expectations. For example,  
observations would suggest bimodal DM profiles, i.e.  deviations from the  predicted 
single power law  distribution in the core of the DM  clumps 
(e.g. \cite{FLO,BBG}). Even assuming  
that there   are at least two distinct mass components (halo and disk) that are contributing to the 
Newtonian gravitational potential \cite{BSL}, a single power law model may be recovered 
but at the cost of a significantly  flatter  slope  ($\beta \sim 0.6 $) than predicted by simulations;\\
{\it c)} the scale below which  our proposed modification of the Newtonian  DM-baryon interaction 
becomes effective is less robustly constrained. While the data of \cite{MAR,BL2} and \cite{BLA} seem to point, 
consistently,  to a value $\lambda = 1.1 \pm 0.08 $ kpc 
(assuming $H_0 = 70$ km s$^{-1}$ Mpc$^{-1}$), the rotation curve of NGC4605 (\cite{BSL}) favors 
a smaller value ($\lambda \sim 0.2\pm 0.02$) kpc. Future velocity surveys of the inner regions of local dwarf
 galaxies will help  in better constraining the range of variation of this fundamental length and in 
understanding if, as we assume  here, NGC 4605 may be considered as non representative of the general 
kinematics of dwarf  galaxies. Finally we note that above  this kiloparsec scale, in the external halo regions, 
a standard Newtonian regime is recovered, and a  NFW density well describe the observed  flattening of the  
rotation curves.

We have also investigated how this  proposed universal form for the coupling of   the DM and baryons     
may help in interpreting  some  CDM results.
Many authors (e.g \cite{WHI}) have argued  that  in order to match  CDM models with observations, 
some mechanism must prevent a large number of dwarf galaxies from forming in low mass halos.
Much attention has been focused on investigating astrophysical and hydrodynamical  mechanisms that 
may suppress the efficiency of converting  baryons into stars in small galactic systems (e.g. \cite{DEK}): 
in particular  supernova-driven winds that expel a large fraction of the 
baryonic component from a dwarf galaxy (e.g. \cite{EFS}, but see \cite{MCF}), and photoionization effects 
that prevent  baryons to cool into stars (e.g. \cite{BEN}). 

\begin{figure}
 \includegraphics[scale=0.54]{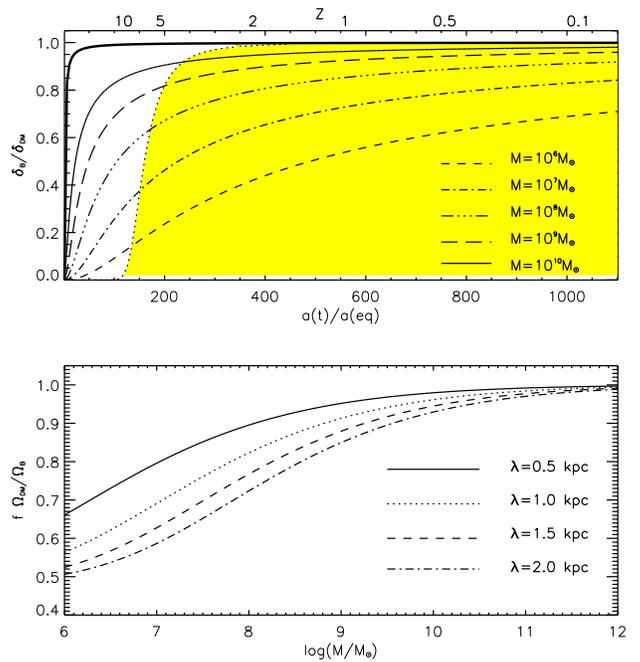}%
 \caption{ 
{\em Upper:} time evolution
of the fraction $\delta_B/\delta_D$ predicted by  our model ($\alpha=-1, \lambda=1.1$ kpc)
for perturbations corresponding to galaxies of different masses. The  
scale-invariant standard result is also plotted (thick line).
The shaded region represents  the post-turn-around epoch. {\em Lower:} Baryon fraction at the turn-around 
epoch   in units of its mean cosmological value $\Omega_B/\Omega_D$) as a function of the mass
of the collapsed object and for different values of $\lambda$. 
\label{due}}
\end{figure}

In our  picture, the lack of a significant luminous component in small halos  may be interpreted noting that,  
in addiction   to local,  baryonic {\it physical} processes, a  {\it cosmological} mechanism  contributes to 
regulate the quantity  of baryons that falls into halos of different masses. 
 The growing of the baryonic density contrast $\delta_B = \delta \rho_B / \bar{\rho}_B $ after decoupling is described 
in linear perturbation theory, neglecting radiation effects \cite{GNE,PM2},
by the equation 
\begin{equation} \label{9}
\ddot{\delta}_B + 2 H \dot{\delta}_B ~=~  \nabla^2 \delta \phi , 
\end{equation} 
where a dot means differentiation with respect to the proper time $t$, $H=\dot{a}/a$ is the hubble parameter and 
$\delta \phi$ is the perturbation of the gravitational potential. DM 
perturbations, approximately the only 
gravity sources, grow almost
independently $\delta_D \equiv \delta \rho_D/\bar{\rho}_D \simeq (a/a_{dec}) \delta_D^{dec}$ 
and create the potential wells where baryons fall in. 
In the standard scenario  $\nabla^2 \delta \phi \simeq 4\pi G\, \delta \rho_D $, and the RHS 
term of \eqref{9} very efficiently drives the growth of baryonic perturbations
of any length scale (the fraction  $\delta_B / \delta_D ~\simeq~ 1 - a_{dec}/a$, in fact,
becomes of order unity already at redshift, say, $z=100$).
By using  \eqref{11} and \eqref{22}, on the contrary, the driving term on the RHS of \eqref{9}  becomes 
scale-dependent: $\nabla^2 \delta \phi ~=~ 4 \pi G (1 - \lambda^2 \nabla^2)^{-1} \delta \rho_D$.
We can gain  some insight  by noting that,  
as long as the proper size $a/k$ of a perturbation keeps small in units of $\lambda \sim$ kpc,
this source can be approximated as
\begin{equation*}
 \nabla^2 \delta \phi ~\approx~  4 \pi G 
\frac{(a/k)^2}{\lambda^2}  \, \delta \rho_D.
\end{equation*} 
A galaxy of mass 
$10^9 $M$_\odot$, for instance, grew out from perturbations of length  $\sim 10^{-1}$ kpc at decoupling; in this case
the relative driving term on the RHS of \eqref{9} is initially suppressed by a factor $\sim 10^{-2}$. 
This dampening is of order $\sim 10^{-4}$ for galaxies of $10^6 $M$_\odot$. 
The full solution of \eqref{9} in our gravitational  scenario is 
plotted in the upper panel of Fig. 2 in terms of the fraction $\delta_B/\delta_D$ and for different masses.

Using this result, it is straightforward to
compute the baryon fraction $f = (\rho_B/\rho_D)$ inside shells reaching  the turn-around. In doing this
we use a $n=1$ power  spectrum  with $\Gamma = 0.25$, $h = 0.7$ and $\sigma_8 = 0.84$
and we implicitly assume that linear theory still holds at this epoch 
(at least for its predictions on the ratio $\delta_B/\delta_D$) and that 
the derived $f$ is a good estimate for the baryon fraction  in  the final  collapsed object.
From the lower panel in Fig. 2 one may see that the Yukawa-like potential acts in such a way to avoid that 
the universal gas fraction ($\Omega_B/\Omega_D$)  be  locked up in low mass clumps at high redshift. For
example,  we predict  $f\sim 0.5 \Omega_B/\Omega_D$  in galaxies having  $M=10^6 M_{\odot}$.
We  note, however,  that to fully exploit the simplicity of linearized equations, we have neglected 
the statistic of the  halo hierarchical merging.

Our phenomenological assumption lacks an explanation in terms of fundamental physics.
The large absolute value of $\alpha$ tends to rule out the interpretation of 
the Yukawa contribution in terms of a scalar field--mediated force. In a scalar-field scenario,
in fact, the strength parameter evaluates  $\alpha = \alpha_B \alpha_D$,
where $\alpha_B$ and $\alpha_D$ are the couplings of the scalar to baryonic and dark matter respectively 
(see e.g. \cite{DPV2}). Since the scalar-baryon coupling is constrained to be 
$|\alpha_B| \lesssim 10^{-2}$ by  
very long baseline radio interferometry measurements (see e.g. \cite{EUB}), a dramatic short-range
modification of gravity of the order of $\alpha_D^2 \gtrsim 10^4$
in the pure dark matter sector could not be avoided,
at least under the minimal assumption of a universal scalar-dark  coupling $\alpha_D$.

In summary,  we have argued  that the coupling between DM and baryons 
vanishes in the small distance limit and it can be effectively described by adding 
a Yukawa-like contribution  with universal parameters
$\alpha \sim -1, \lambda \sim 1 $ kpc to the
standard gravitational potential. 
Our model does not affect the relation between dark and visible matter on \emph{large} cosmological scales 
(i.e. the dynamics of large galaxies and clusters) but interestingly
it successfully alleviates  some inconsistencies between observations and 
CDM theoretical  predictions on kpc scales. In particular: \\
a)  the analysis of the H$_\alpha$ rotation curves of LSB galaxies  shows that the 
central dark matter density distribution  predicted under the assumption of a Yukawa potential well 
($\beta =-1.35\pm 0.05$) is in excellent agreement with the cuspy profile derived in collisionless 
CDM simulations.
In particular, the $\alpha$ best fitting values suggest that, independently of the 
adopted Yukawa model, the gravitational interaction between these two 
classes of matter is totally suppressed in the small distance limit
b)  this potential offers an interesting  cosmological mechanism to segregate the quantity of baryons
that falls into primordial  dark matter perturbations as a function of the fluctuation mass; this fact 
may be used to relax   estimates of the efficiency of various feedback mechanisms proposed to 
suppress the overproduction  of dwarf galaxies.

We acknowledge precious discussions with  A. Bosma, B. Bassett, M. Bruni, 
R. Giovanelli
and F. Vernizzi. FP thanks the Laboratoire d'Astrophysique de Marseille for hospitality.
CM acknowledges financial support from the Centre National de la Recherche Scientifique and the Region PACA.

\end{document}